\documentclass[a4paper,11pt]{article}
\pdfoutput=1 

\usepackage{jinstpub} 
\usepackage[utf8]{inputenc}

\title{About the possibility of an international ground-based Very High Energy particle detector experiment in Ecuador}

\author[a,1]{D. Cazar Ramirez,\note{Corresponding author.}}
\author[b]{G. Violini,}

\affiliation[a]{Colegio de Ciencias e Ingenierías "El Politécnico" \\ Universidad San Francisco de Quito USFQ,\\Quito, Ecuador}
\affiliation[b]{Centro Internacional de Física CIF,\\Bogotá, Colombia}

\emailAdd{dcazar@usfq.edu.ec}

\abstract{
We discuss the possibility of hosting a big Astrophysics ground-based experiment in Ecuador aimed to detect VHE particles. Ecuador location makes possible to see both the Northern and Southern sky. An additional geographic feature is the presence of one of the highest American mountains, the Chimborazo (6310 m.a.s.l.), that happen to be the highest point on Earth measured from the center of the planet. In the last decade Ecuadorian government has invested resources in higher education and research, with an important policy of training abroad. The effect has been that now many researchers in Physics, Astrophysics, and Engineering are working in universities across the country and collaborating with important experiments like CMS at CERN, Pierre Auger in Argentina, HAWK in Mexico and in LAGO project. All these features make Ecuador an ideal place to host a big VHE particle detector experiment in South America.

}


\proceeding{XIV  Workshop on Resistive Plate Chambers and Related Detectors\\
  19-23 February 2018 \\
  Mexico}

\begin{document}
\maketitle
\flushbottom

\section{Contents}
\label{sec:intro}
Very High Energy Cosmic Rays (VHECR) research has various interesting questions open about their origin and composition, their interaction with our planet and the effects on human beings \cite{a}. The study of this phenomena brings new insights in fundamental physics and boost the development of state of the art technology in electronic systems, computer simulation techniques, material science among others. Applications covers biophysics (prediction and monitoring of CR dosis in astronauts) and medicine (cancer treatment using charged particles). Direct observations of CR is still the only way to study VHECR (above $10^{13}eV)$ since elementary particle at those levels of energy cannot be “produced” in laboratory. Satellite, airborne and ground based experiments have been developed in the past years around the world.
This article is focused in VHECR experiments at ground level using different approaches and techniques \cite{b} that are located in the southern hemisphere. After a brief description of the experiments it follows a description of Ecuador as a potential candidate to host a VHECR experiment, identifying some locations at high altitude, and finalizing with a study of the impact that a VHECR experiment would generate in the development of science and scientific culture in Ecuador and the Latin America.

\section{Cosmic ray detection experiments in Latin America}
The most important experiments in cosmic ray detection in Latin America are Pierre Auger Observatory, Chacaltaya Cosmic Ray Laboratory and LAGO Project. A brief description of each one follows.

\subsection{Pierre Auger Observatory}
Based in relays on an hybrid detection technique (surface detectors plus fluorescence telescopes) to detect EeV photon fluxes, a huge collecting area of about 3000 $km^2$ where 1660 Water Cherenkov Detectors (WCDs) are installed allows to obtain the secondary particle density at ground level; the 27 telescopes installed to detect the air fluorescence are used to measure the longitudinal development of the air shower about the surface level \cite{c}. Based in Malargüe (Mendoza, Argentina) at 1402 m.a.s.l. Pierre Auger Observatory can reconstruct with high accuracy the direction of the primary cosmic ray which provoked the secondary particle shower. Therefore the source of the cosmic ray can be identified, composition of the shower can be studied as well since WCDs detects charged particles, mostly muons, and electrons.
Pierre Auger Observatory is operated by the Pierre Auger Collaboration, a consortium of research institutions from 16 countries. In Argentina there are 7 institutions participating and collaborating in the project. All of these institutions have developed strong research groups in
astrophysics, high energy physics, particle detection and space weather. To be involved in an experiment as Pierre Auger is certainly a benefit in the long run. An extensive education and outreach program has been carried out in Malargüe region \cite{d} with interesting results, tourists and locals can visit the Auger Visitor center and conferences and seminars are organized in elementary and high schools nearby. All this activities increase the interest of young students and local population about basic sciences and engineering.

\subsection{Chacaltaya Cosmic Ray Laboratory}
The facilities are located at the top of the Mount Chacaltaya near La Paz Bolivia, at 5220 m.a.s.l. Due to its altitude a low energy threshold (13.1 GeV) can be reached \cite{e} making this location an ideal site for study of $\gamma$-astronomy. Several experiments are running in Chacaltaya operated and founded by institutions as the Centro Latinoamericano de Física and Brazilian and Japanese groups. This is the oldest high altitude CR laboratory in the world, from those installations important contributions to astrophysics has been done. The Chacaltaya Observatory hosts a variety of other experiments as the Global Atmosphere Watch, an initiative of the World Meteorological Organization, monitoring meteorological variables, aerosols and greenhouse gases \cite{f}.

\subsection{The LAGO Project}
The Latin American Giant Observatory is an extended cosmic ray observatory composed by a network of WCDs spanning over different sites located at significantly different altitudes (from sea level up to more than 5000 m.a.s.l.) and latitudes across Latin America, covering a huge range of geomagnetic rigidity cut-offs and atmospheric absorption/reaction levels \cite{g}. This detection network is designed to measure the temporal evolution of the radiation flux at ground level with extreme detail. The LAGO project is mainly oriented to perform basic research in three branches: high energy phenomena, space weather and atmospheric radiation at ground level. LAGO is built and operated by the LAGO Collaboration, a non-centralized collaborative union of more than 30 institutions from ten countries \cite{h}. Through LAGO a big and strong community of latin american scientists has been
created and along with the scientific contributions made, training of young scientists and students of all over the continent it is a major activity of the collaboration. In Ecuador, three of the most prestigious Universities are members of LAGO, ecuadorian research groups work in computer simulations, data acquisition systems developing and data analysis.

\section{Description of Ecuador}
\subsection{Geographic description}
Ecuador is located between latitudes $2^oN$ and $5^oS$. The Cordillera de los Andes mountains chain crosses the country from north to south creating a highland region with high altitude plateaus and mountains and volcanoes. Ecuador location makes possible to see both the Northern and Southern sky. Three sites can be considered as possible candidates to host the experiment:

\begin{itemize}
    \item Chimborazo is the highest mountain in Ecuador (6310  m.a.s.l.), it is located 30 km away from Riobamba and 180 km from Quito. Some plateaus above 4000 m.a.s.l. has been identified as an ideal location for VHECR experiments \cite{i}.
    \item Cotopaxi is the second highest mountain in Ecuador (5897 m.a.s.l.) is located 50 km south of Quito, high altitude tundras nearby are easy to access. A former NASA satellite tracking station at 4200 m.a.s.l. is now a base of the Instituto Espacial Ecuatoriano (IEE) and can be used a basecamp.
    \item El Cajas National Park (3100 to 4450 m.a.s.l.) in the souther region of Ecuador is located 30 km south of Cuenca, it is a high plateau with easy access.
\end{itemize}

\subsection{Education and research facilities}
Solid research groups in astrophysics, gamma ray burst and scientific instrumentation had grown in Universities located in Quito, Universidad San Francisco de Quito (USFQ) and Escuela Politécnica Nacional (EPN), and Riobamba (Escuela Superior Politécnica de Chimborazo (ESPOCH). These institutions are members of LAGO Collaboration since 2010, all of them have physics and electronics careers.
A new research group of Universities from Cuenca, a city near El Cajas National Park ( 3100 to 4450masl) in the southern region of Ecuador, have started to collaborate with LAGO. Universidad del Azuay and Universidad de Cuenca are planning to build a WCD detector and placed at El Cajas. It has to be noticed that neither of those institutions have a physics career even though Universidad de Cuenca is one of the oldest educational institutions in Ecuador. The Corporación Ecuatoriana para el Desarrollo de la Investigación y la Academia (CEDIA) \cite{j} is a non-government consortium of universities created to generate and strengthen research networks, both nationally and internationally. CEDIA provides financial support for research and development projects through the CEPRA program and makes available to the Universities a wide variety of services as high speed internet access, high performance equipment for data processing (HPCs and data repositories), free access to licenced software (Wolfram technologies for education). In preliminary conversations with CEDIA management board a high interest into participate in a VHERC experiment has been manifested. The Secretaria de Educación Superior, Ciencia, Tecnología e Innovación Senescyt \cite{k} is a government institution aiming to promote the development of research activities and technology development through a very aggressive fellowship program for students to join master and PhD programs around the world plus annual contests for founding research projects. Recently
Senescyt financed the inclusion of Ecuador as member of the CMS Collaboration at CERN confirming the interest of the ecuadorian government in the development of particle physics.

\subsection{International research centers}
One of the authors has been pushing hard to create a research center (of Unesco category 2) supported by ICTP inspired in the style of those of Chiapas and Sao Paolo. The main purpose of the center would be to provide a link between European research institutions connected to ICTP
and Andean region research institutions \cite{l}. The project proposed two possibilities about the organization and scope of the center:

\begin{itemize}
    \item To develop specific branches of science and technology based on the experience and human resources of the possible beneficiary institutions of the region, research in astrophysics and related topics would be a natural choice.
    \item To develop a general research center in basic and applied science, taking into account the present condition of the region this choice is less interesting \cite{m}.
\end{itemize}

Thinking about the future a project like this can easily cover other regions in Latin America
where developing countries as caribbeans will beneficiate participating in the activities that such
research center can provide and organize \cite{n}.

\section{A VHECR international experiment in Ecuador}

The main idea is to build a high altitude ground based VHECR observatory in one of the locations listed above. There exists the possibility to retrieve the equipment installed in Tibet for Argo-JBY observatory \cite{o}. Ecuadorian government has to activate all the diplomatic channels needed to make this happen. \\
Argo-YBJ uses Resistive Plate Chambers RPCs as detectors, a novel technique that ensure high efficiency and low cost of production, relatively easy installation and operation. A project to update the technology, gas production with low environmental impact and the development of new data acquisition systems can be a solid starting point for researchers and engineers to develop and construct functional prototypes to validate designs and increase the knowledge about the technique. This activities can be carried out while the Argo-YBJ equipment arrives in Ecuador and the building blocks in the chosen sites are finished.\\
The creation of and Information center and outreach activities program as in Pierre Auger Observatory is to consider. This activity will bring benefits in the long term. The center can function as both a tourist attraction and a place to hold conferences for dissemination of
scientific knowledge to motivate students from elementary and high schools of the country into study fundamental sciences and eventually be a part of the experiment, Corporación Ecuatoriana para el Desarrollo de la Investigación y la Academia CEDIA is very interested in to provide the technology and the knowledge they have to build a maintain a high speed communication network for data transmission from the experiment site to
the nearest research facility, provide HPC resources to run simulations and modelling of the detectors and data storage.\\
Open the possibility to other countries from Latin America and Europe to participate in order to enhance the scientific and technological knowledge and to ensure financial support for the time the experiment will operate.

\section{Conclusions}

\begin{itemize}
    \item Ecuador fulfills the basic characteristics that a country should have to host a high altitude VHECR experiment. High altitude plateaus are near to cities where research facilities can be built simplifying the logistic and lowering the cost to build constructions near the sites.
    \item Universities and research institutions have both the human and the technological resources needed to support the developing phase of the project. Particle detection laboratories already exists, a HEPCR experiments will help them to grow and expand.
    \item Researches and engineers in Ecuador are members of international collaborations allowing them to learn the way of working in a multilevel, multiphase project as a VHECR experiment is.
    \item Government and private institutions as Senescyt and CEDIA may be involved in the project providing important economic resources for specialization, researchers mobility, high speed connectivity, HPC resources and data repositories.
    \item Outreach activities directly related to a VHERC experiment will have a positive impact in the development of tourism, commerce as well as motivate students near the site (elementary and high school) to study fundamental science and increase the interest of the society to research topics.
    \item Creation of a regional research center with ICTP support can boost the development of science and technology in the andean region and a connection with the experiment would be a plus. To train researchers and engineers in RPCs particle detection techniques may start the development of applications in other fields rather than VHEP detection. Due to the abundance of active volcanoes in the regions, muon tomography may be explored and applications. Applications of RPCs in medical physics may be explored as well.
\end{itemize}

\acknowledgments

This research was partially supported by Red Nacional de Investigación y Educación del Ecuador RedCEDIA through the grant CEPRAXI-2017-Astroparticulas-2.\\
We thank professor Rinaldo Santonico from Universitá di TorVergata (Rome, Italy) who provided insight and expertise that help us to develop the main idea, although they may not agree with all of the interpretations/conclusions of this paper. \\
We also thank Edgar Carrera from Universidad San Francisco de Quito for supporting the idea and the time expended revising this work.



\end{document}